\documentclass[aps,prb,twocolumn,groupedaddress]{revtex4-1}
\usepackage[latin9]{inputenc}
\usepackage{graphicx}
\usepackage{epstopdf}
\makeatletter
 
 \@ifundefined{textcolor}{}
 {%
   \definecolor{BLACK}{gray}{0}
   \definecolor{WHITE}{gray}{1}
   \definecolor{RED}{rgb}{1,0,0}
   \definecolor{GREEN}{rgb}{0,1,0}
   \definecolor{BLUE}{rgb}{0,0,1}
   \definecolor{CYAN}{cmyk}{1,0,0,0}
   \definecolor{MAGENTA}{cmyk}{0,1,0,0}
   \definecolor{YELLOW}{cmyk}{0,0,1,0}
 }

\usepackage{dcolumn}
\usepackage{bm}
\usepackage{xcolor}

\makeatother

\begin{document}






\title{Two-phonon Raman bands of single-walled carbon nanotubes: a case study}



\author{Valentin N. Popov}
\affiliation{Faculty of Physics, University of Sofia, BG-1164 Sofia, Bulgaria}


\date{\today}
\begin{abstract}
It has been long accepted that the second-order Raman bands in carbon nanotubes are enhanced through the double-resonance mechanism. Although separate aspects of this mechanism have been studied for a few second-order Raman bands, including the most intense defect-induced D band and the two-phonon 2D band, a complete computational approach to the second-order bands is still lacking. Here, we propose such an approach, entirely based on a symmetry-adapted non-orthogonal tight-binding model with ab-initio-derived parameters. As a case study, we consider nanotube $(6,5)$, for which we calculate the two-phonon spectrum. We investigate in detail the 2D band and identify three contributions to it: a non-dispersive one and two dispersive ones, which are found to depend on the electron and phonon dispersion of the nanotube, and on the laser excitation. We also predict two-phonon bands, which are not allowed in the parent structure graphene. The obtained two-phonon bands are in very good agreement with the available experimental data. The symmetry-adapted formalism makes feasible the calculation of the two-phonon Raman bands of any observable nanotube.  
\end{abstract}
\maketitle


\section{Introduction}

In the last few decades, the layered carbon materials like fullerenes, nanotubes, and few-layer graphene have been a subject of intense experimental and theoretical study, because of their unique properties, originating from their zero-, one-, two-, and three-dimensionality.\cite{dres10} In particular, significant progress has already been achieved in the synthesis and the study of the properties and application of carbon nanotubes\cite{rtm04,jori08} The application of nanotubes in nanoelectronics requires their precise structural characterization. For this purpose, the Raman scattering by phonons is the experimental technique of choice, being a fast and nondestructive characterization method.\cite{thom07} 

The so-called single-walled nanotube (or briefly \textit{nanotube}) can be viewed as obtained by rolling up a graphene sheet into a seamless cylinder. It has a few intense first-order Raman bands, arising from the radial-breathing mode (\textit{RBM}) and the longitudinal and transverse tangential modes (\textit{G modes}). The RBM frequency is found to be inversely-proportional to the nanotube radius and is normally used for fast sample characterization.\cite{arau08} The G modes also depend on the nanotube radius and can be used to support the assignment of the Raman spectra to particular nanotubes but also allow for differentiating between metallic and semiconducting nanotubes.\cite{jori02} Intense second-order Raman bands are also observed in nanotubes. These bands arise from scattering processes, which can involve two phonons with opposite momenta (\textit{two-phonon bands}) or  a phonon and a defect (\textit{defect-induced bands}). The most intense two-phonon band (so-called \textit{2D band}), usually observed around $2700$ cm$^{-1}$, is due to electron/hole scattering by two transverse optical (TO) phonons with opposite momenta. Other two-phonon bands have also been measured (for summary, see, e.g., Ref.~\onlinecite{tybo18}). The two-phonon spectra contain valuable information about the phonon dispersion of the nanotubes.\cite{maul03}

The D and 2D band intensity is found to be enhanced for certain laser excitations, which has been explained by the the double-resonant (DR) scattering mechanism, initially proposed for graphene.\cite{thom00} In the early days of the modeling of these bands in graphene, simplifying assumptions such as constant electronic linewidth, limiting the calculations only to high-symmetry directions in the Brillouin zone, and exact DR conditions have been used (for a review, see, e.g., Ref.~\onlinecite{vene11}). Only recently, theoretical two-phonon bands in graphene, which are in good agreement with the experimental data, have been reported.\cite{vene11,popo12} The success of the latter studies is based on the explicit calculation of the couplings and the electronic linewidth, and performing the integration of the quantum-mechanical expression for the Raman intensity over the entire Brillouin zone for both electrons and phonons.  

The theoretical description of the two-phonon bands of nanotubes is much more complex than for graphene, mainly, because of the quantum confinement in one dimension and the large variety of nanotube types. Recently, a step towards the understanding of the defect-induced D mode in nanotubes vs diameter and energy has been made using the hexagonal symmetry of graphene and geometrical considerations.\cite{herz15,vier17} This approach allows assigning the two-phonon bands to pairs of definite phonons. However, the two-phonon band shape can only be predicted, taking into account the couplings and the electronic linewidth, and carrying out full integration over the Brillouin zone. In a recent publication,\cite{mour17} this has been done for a few narrow nanotubes using ab-initio and tight-binding approaches, neglecting, however, the effects of the nanotube curvature.  

Here, we propose a computational approach to the calculation of the two-phonon Raman spectra of carbon nanotubes, which is entirely based on a symmetry-adapted ab-initio-based non-orthogonal tight-binding (NTB) model.  This model has been used for more than a decade for the successful prediction of the electronic band structure and phonon dispersion, and the first-order Raman bands of several hundred nanotube types.\cite{popo04,popo05,popo06,popo06a,popo10a} The model describes the curvature effects on the physical properties of carbon nanotubes, which are essential for nanotube diameters below about $1$ nm. As a case study, we consider the narrow nanotube $(6,5)$ and perform a complete calculation of the two-phonon bands at a number of laser excitations and discuss the contributions to the 2D band, as well as the appearance of two-phonon bands, which are symmetry-forbidden for the parent structure graphene.
 
The paper is organized as follows. The theoretical background is presented in Sec. II. The obtained results are discussed in Sec. III. The paper ends up with conclusions, Sec. IV. 

\section{Theoretical background}

A nanotube can be considered as obtained by cutting out a rectangle of graphene, defined by a pair of orthogonal lattice vectors $\vec T $ and $\vec C$, and rolling it along $\vec C$ into a seamless cylinder. This rolled-up nanotube can be characterized by the radius $R = \| \vec{C} \|/2\pi$, translation period $T \equiv \| \vec{T}\|$, as well as by the chiral angle $\theta$, which is the angle between $\vec C$ and the nearest zigzag of carbon atoms. All structural parameters of the rolled-up nanotube can be expressed by means of the nearest-neighbor interatomic distance and the hexagonal indices $(n,m)$ of $\vec{C}$ or $(\tilde{n},\tilde{m})$ of $\vec{T}$. The former notation is traditionally used to specify uniquely the nanotube and is accepted here as well. Normally, the total energy of the rolled-up nanotube is not minimal and the atomic structure of the nanotube has to be subjected to structural relaxation, which is a necessary step before performing phonon calculations. 

The straightforward calculation of the electronic states and phonons for a large variety of nanotubes is accompanied by insurmountable computational difficulties, because of the very large translational unit cells of most of the observed nanotubes. Fortunately, the nanotubes have screw symmetry that allows reducing the computational efforts by resorting to two-atom unit cells. The latter approach has been used for calculation of the electronic states\cite{popo04} and phonons\cite{popo06} of several hundred nanotubes within the NTB model. In this model, the Hamiltonian and overlap matrix elements are derived as a function of the interatomic separation from an ab-initio study on carbon dimers\cite{pore95} and the Slater-Koster scheme is adopted for the angular dependence of the matrix elements. 

The imposing of the translational periodicity and rotational boundary conditions on the solutions of the electron and phonon eigenvalue problems results in labeling of these solutions by a pair of indices. The electronic states are labeled by the one-dimensional wavevector $k$, $k \in [0,1)$  $2\pi/T$, and the integer quantum number $l$, $l \in [0,N)$, where $N$ is the number of two-atom unit cells in the translational unit cell of the nanotube. Similarly, the phonons are labeled by the one-dimensional wavevector $q$, $q \in [0,1)$  $2\pi/T$, and the integer quantum number $\lambda$, $\lambda \in [0,N)$. The selection rules for scattering of electrons by phonons are $k' = k + q \pm 2\pi/T  $ and $l' = l + \lambda \pm L$; $L = (N\nu +n)/\tilde{n}$, where $\nu$ is an integer number such that $L$ is an integer number. These selection rules can be derived similarly to those in Ref.\onlinecite{popo06a} by including Umklapp processes. Notice that the integer quantum number is non-conserving.\cite{vuko02}

In quantum mechanics, the two-phonon process can be viewed as a sequence of virtual processes, namely, electron-hole creation, scattering of electrons or holes by two phonons with opposite momenta, and finally electron-hole annihilation. The wavevector is conserved for each virtual process, while the energy is conserved only for the entire two-phonon process. The corresponding two-phonon Raman intensity can be described by an expression, derived in fourth-order quantum-mechanical perturbation theory,\cite{mart83,vene11,popo12,thom07,jori08}

\begin{equation}
I\propto\sum_{f}\left|\sum_{c,b,a}\frac{M_{fc}M_{cb}M_{ba}M_{ai}}{\Delta E_{ic}\Delta E_{ib}\Delta E_{ia}}\right|^{2}\delta\left(E_{i}-E_{f}\right)\label{a2}
\end{equation}

Here, 
$\Delta E_{iu}=E_{iu}-i\gamma$ and $E_{iu} = E_{i}-E_{u}$; $E_i$ is the energy of the initial state; $E_{i}=E_{L}$, where $E_{L}$ is the incident photon energy (laser excitation); $E_{u}$, $u=a,b,c,f$, 
are the energies of the intermediate ($a,b,c$) and final ($f$) states of the system of photons, electrons, holes, and phonons.  $M_{uv}$ are the matrix elements between initial, intermediate, and final states. $M_{ai}$ and $M_{fc}$ are momentum matrix elements.   $M_{ba}$ and $M_{cb}$ are electron-phonon matrix elements. The electron-photon and electron-phonon matrix elements are calculated explicitly within the NTB model.\cite{popo05} $\gamma =\gamma_c + \gamma_v $, where $\gamma_c$ and $\gamma_v$ are the halfwidths of conduction ($c$) and valence ($v$) states, respectively.\cite{popo06a}  The summation over the phonon wavevectors $q$ and the electron wavevector $k$ is carried out over a mesh of $400$ points in the Brillouin zone. The Raman intensity is enhanced for vanishing of one, two, or three of the $E_{iu}$'s in Eq. 1, and therefore single, double, and triple resonances can occur.\cite{vene11} The incident and scattered light are assumed polarized along the nanotube axis and only Stokes processes are considered in this work.

\section{Results and Discussion}

We consider nanotube $(6,5)$ for exemplifying the behavior of the two-phonon bands vs laser excitation. The translational unit cell of the nanotube contains $N = 182$ two-atom unit cells. The atomic structure of the nanotube is relaxed, retaining its circular cylindrical form. In this case, as independent structural parameters, we choose the radius $R$, the translation period $T$, and the coordinates of the second atom relative to the first atom in the zeroth two-atom unit cell, and obtain the relaxed parameters $R =  3.76$ \AA~ and $T = 40.67$ \AA.

\begin{figure}[tbph]
\includegraphics[width=40mm]{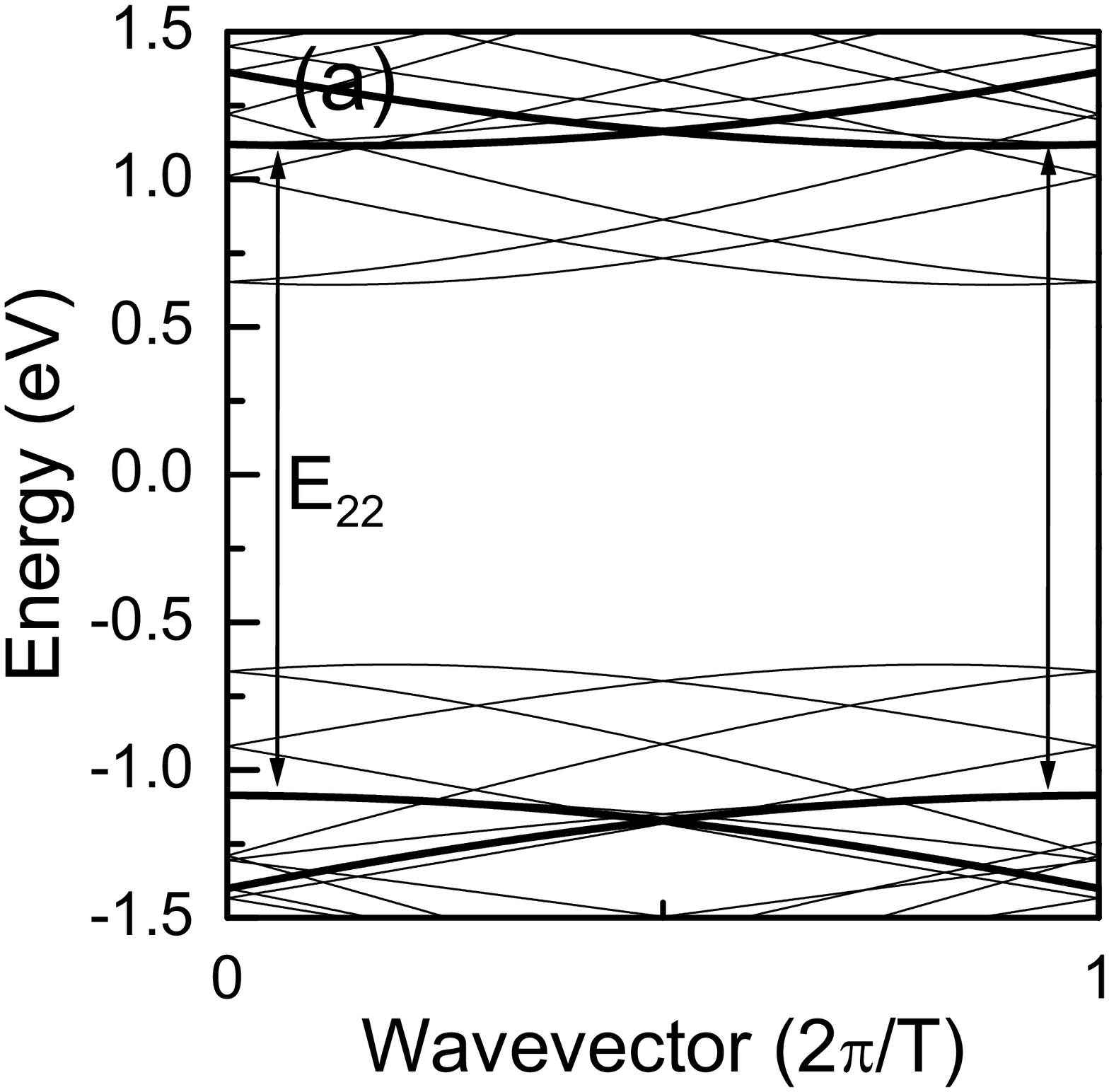} 
\includegraphics[width=40mm]{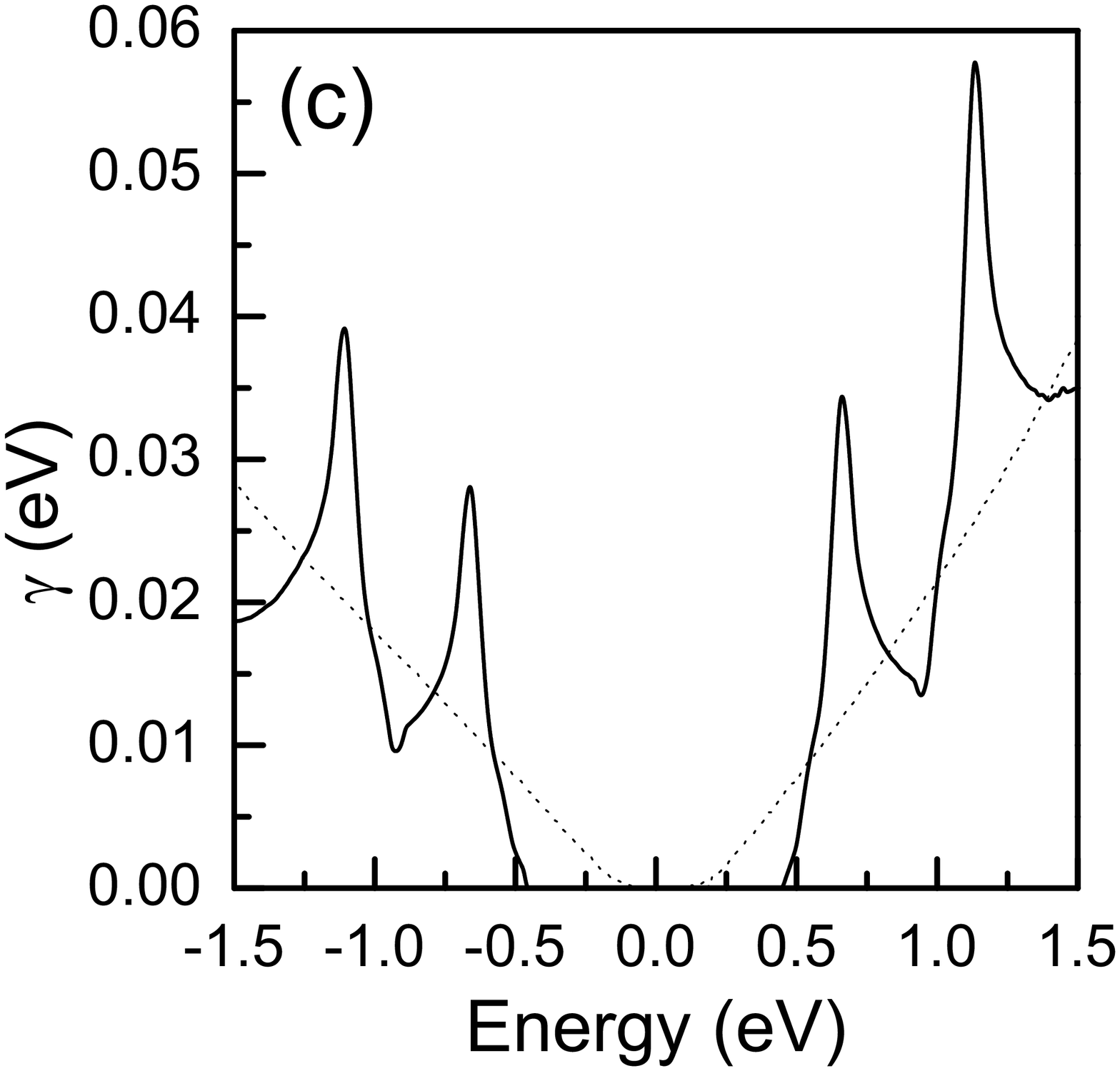}
\includegraphics[width=80mm]{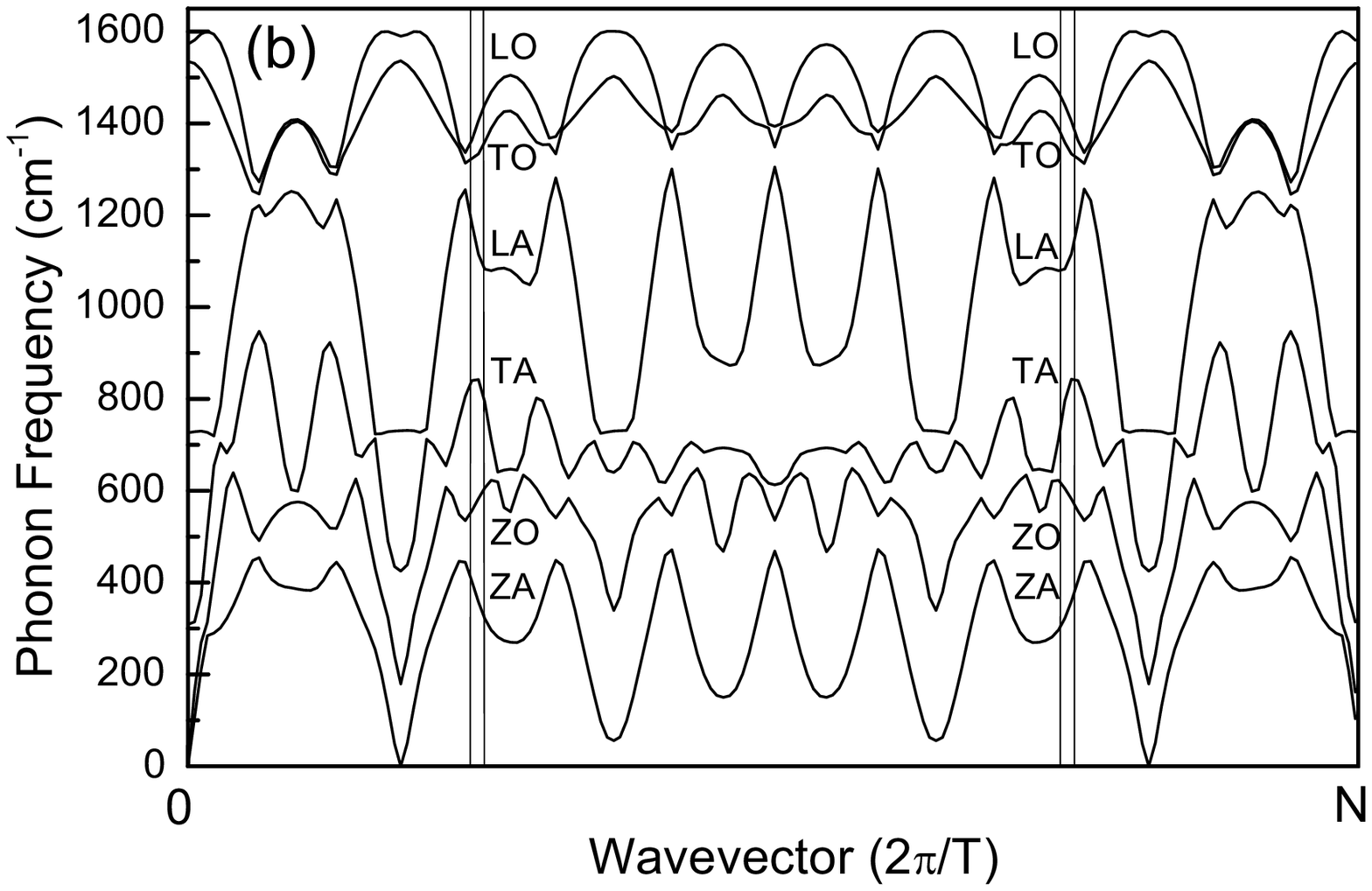}
\caption{(a) Electronic band structure of nanotube $(6,5)$ close to the Fermi energy, chosen as zero. The optical transition $E_{22}$ is denoted by vertical arrows. (b) Phonon dispersion of nanotube $(6,5)$ in the extended zone representation. The pairs of close vertical lines bracket the phonons, taking part in scattering between the extrema of the parabolic bands for the $E_{22}$ transition. (c) Halfwidths of valence ($\gamma_v$) and conduction ($\gamma_c$) states of nanotube $(6,5)$. The corresponding halfwidths for graphene (dotted lines) are provided for comparison.}
\end{figure}

The electronic band structure of nanotube $(6,5)$ in the entire Brillouin zone and for energies up to $\pm 1.5$  eV relative to the Fermi energy $E_F = 0$ eV in shown in Fig. 1a. The valence and the conduction bands in this energy range are bands of graphene of mainly $\pi$ and $\pi ^{\ast}$ character, folded along the cutting lines of the Brillouin zone of graphene.\cite{sait05} None of the cutting lines passes through the K and K$'$ points of the Brillouin zone of graphene and therefore there are no linear bands that cross the Fermi energy in the Brillouin zone of the nanotube. Thus, unlike graphene, nanotube $(6,5)$ is a semiconductor. Due to the presence of electronic bands with extrema (so-called \textit{parabolic bands}), the electronic density of states for such bands has singularities at the energies of the extrema, so-called van Hove singularities (\textit{vHs}) (not shown). The absorption of electromagnetic radiation is enhanced for photon energies, matching mirror pairs of vHs with respect to the Fermi energy.\cite{popo04} In the mentioned energy range, the NTB model predicts only two optical transitions for nanotube $(6,5)$, which take place between the first and the second pair of mirror vHs and are denoted as $E_{11}$ and $E_{22}$, respectively. Here, we will be interested in the parabolic bands, associated with the transition $E_{22} = 2.20$ eV.

\begin{figure}[tbph]
\includegraphics[width=80mm]{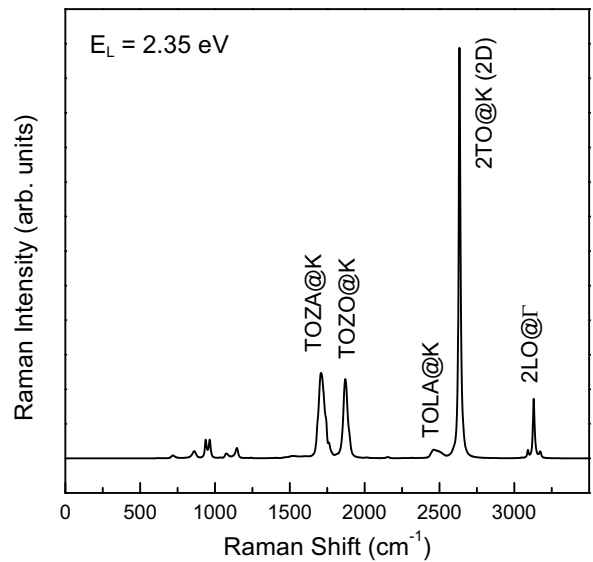} 
\caption{(a) Calculated two-phonon Raman spectrum of nanotube $(6,5)$ at laser excitation $E_L = 2.35$ eV. The most intense two-phonon bands are labelled. The spectrum is slightly upshifted for clarity.}
\end{figure}

The phonon dispersion of nanotube $(6,5)$ consists of $6N$ branches and their representation in the Brillouin zone for the translational unit cell is not very informative, because the branches would cover densely the graph. Alternatively, one can draw fewer phonon branches in the extended zone representation,\cite{doba03} namely, $6\nu$, where $\nu$ is the greatest common divisor of $n$ and $m$. In the case of nanotube $(6,5)$, $\nu =1$ and therefore there are only $6$ branches in this representation. These branches have in-plane longitudinal optical (LO), transverse optical (TO), longitudinal acoustic (LA), and transverse acoustic (TA) character, and out-of-plane optical (ZO) and acoustic (ZA) character. The unfolded phonon dispersion consists of parts of width $2\pi /T$ characterized by different values of the integer quantum number (Fig. 1b). The selection rules for the wavevector and the integer quantum number impose restrictions on the allowed values of the phonon wavevectors $\pm q$ for scattering of electrons between the parabolic bands for transition $E_{22}$ to the regions between the pairs of close vertical lines in Fig. 1b. 

The Raman intensity, Eq. 1, depends crucially on the electronic linewidth. The calculated linewidth for nanotube $(6,5)$ exhibits sharp spikes, arising from the singularities of the electronic density of states, while that for graphene is a smooth function of energy (Fig. 1c). The singularities of the linewidth effectively decrease the Raman intensity close to the transition energies of the nanotube.

\begin{figure}[tbph]
\includegraphics[width=80mm]{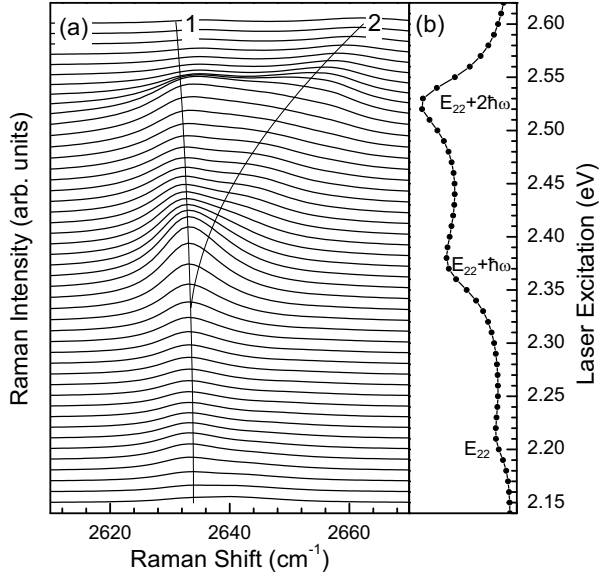} 
\caption{(a) Calculated Raman 2D band of nanotube $(6,5)$ vs laser excitation $E_L$. The 2D band shape evolves from a single- to two-peaked structure (peaks $1$ and $2$) with increasing $E_L$ from $2.15$ to $2.60$ eV with a step of $0.01$ eV (see right panel axis). (b) Integrated Raman intensity of the 2D band (in arb. units) vs $E_L$. Three peaks, centered at $E_L = E_{22}$, $E_{22} + \hbar \omega$, and $E_{22} + 2\hbar \omega$, where $\omega$ is the TO phonon frequency, are clearly seen. The line is a guide to the eye.}
\end{figure}

The derived electronic band structure and phonon dispersion, halfwidths, electron-photon and electron-phonon matrix elements are used in Eq. 1 for calculating the two-phonon Raman spectrum of nanotube $(6,5)$ at $E_L = 2.35$ eV (see Fig. 2). The most intense band in this spectrum is positioned at $\approx 2630$ cm$^{-1}$. This band arises from TO phonons in the vicinity of the K point of the hexagonal graphene Brillouin zone and is usually denoted as 2TO@K or simply as 2D band. The 2D band, along with two less intense bands - TOLA@K at $\approx 2450$ cm$^{-1}$ and  2LO@$\Gamma$ at $\approx 3200$ cm$^{-1}$ - are observed in the Raman spectra of the parent structure graphene. In nanotubes, bands around $1950$ cm$^{-1}$ have been observed and assigned to scattering processes TOLA@$\Gamma$ or LOLA@$\Gamma$  (for discussion, see, e.g., Ref.~\onlinecite{tybo18}). Here, we find that such combination modes have negligible intensity and are unlikely to be connected to the observed Raman bands. On the basis of our results, we assign such experimental bands to the predicted here combination bands TOZA@K and TOZO@K. Such combination modes are not observed in graphene, because scattering of electrons by ZA and ZO phonons is not allowed. However, these scattering processes become allowed in nanotubes due to their curved surface.\cite{tybo18}

We focus on the behavior of the 2D band vs laser excitation $E_L$ in the range from $2.15$ up to $2.60$ eV (Fig. 3a).  It is clearly seen, that the 2D band undergoes an evolution from a single, Lorentz-like shape at small $E_L$ to a two-peaked structure at large $E_L$. The separation between the two components of the 2D band is practically zero for energies below $2.35$ eV but increases up to a few tens of cm$^{-1}$ at $2.60$ eV.  The integrated Raman intensity exhibits three peaks with separation between the adjacent peaks of about  $\hbar \omega$, where $\omega$ is a characteristic TO phonon frequency (Fig. 3b). We have not attempted fitting the 2D band with two Lorentzians and deriving the Raman excitation profile for each of them, because of the strong overlap of the two peaks of this band.

\begin{figure}[tbph]
\includegraphics[width=40mm]{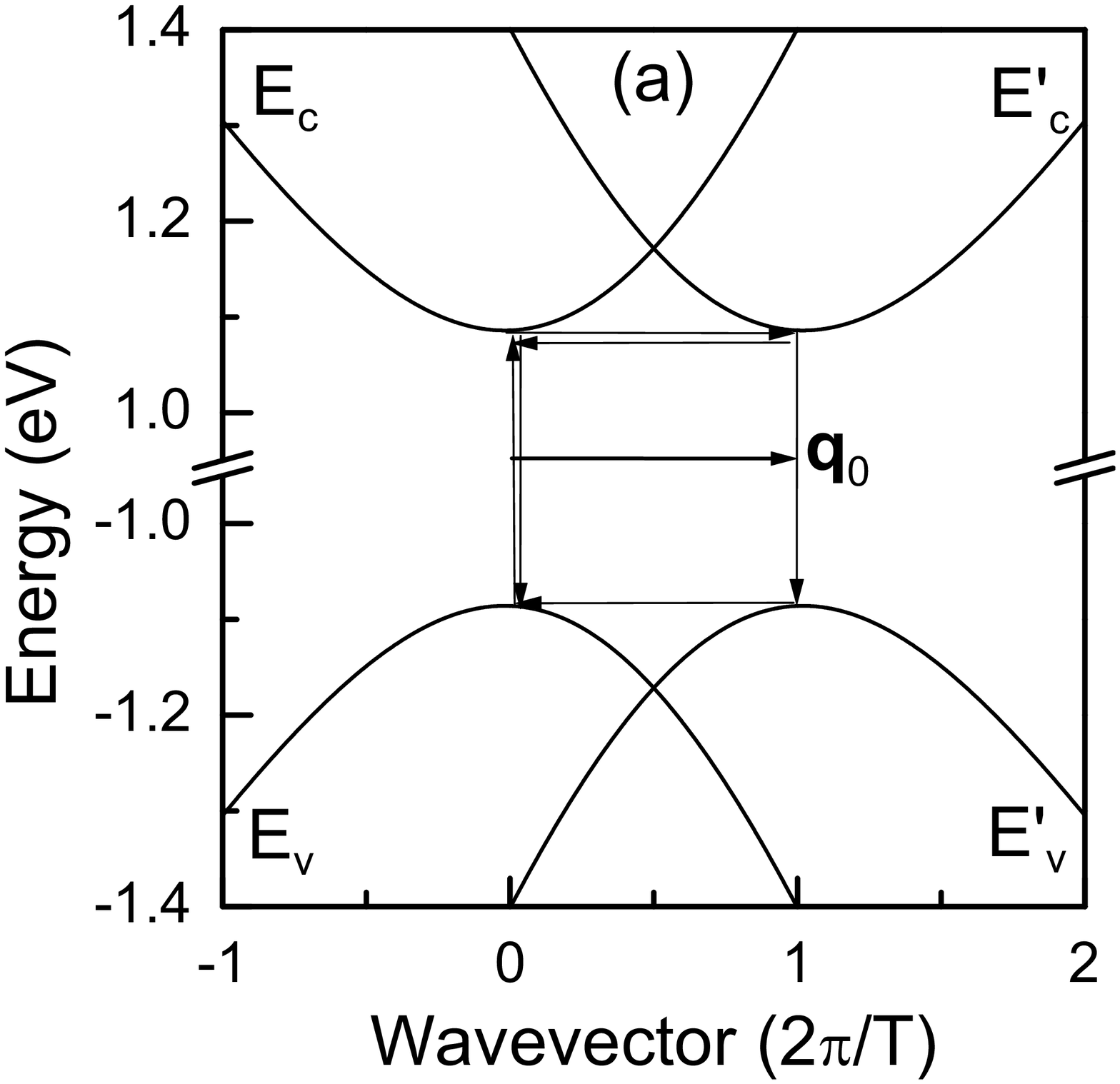} 
\includegraphics[width=40mm]{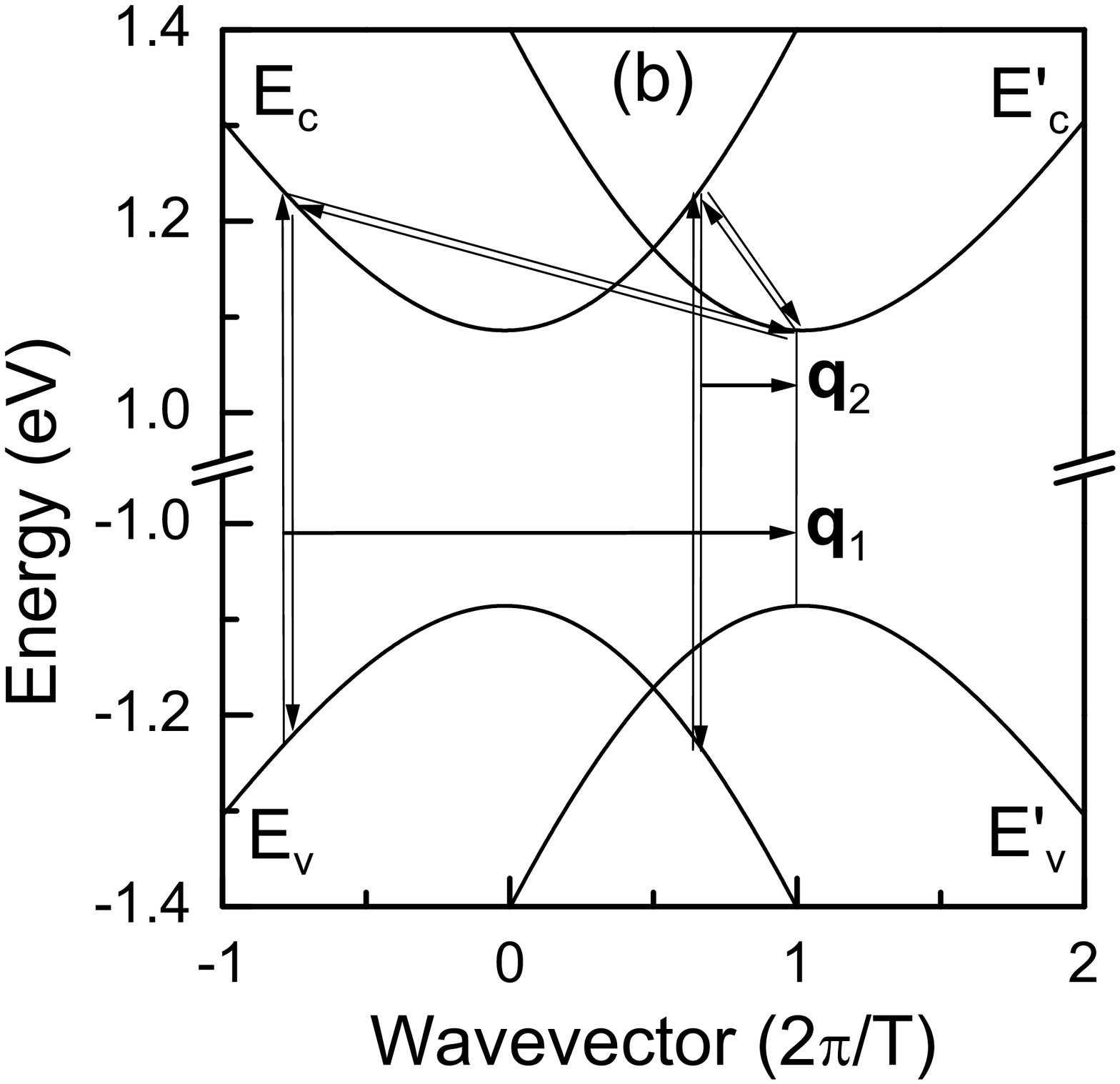} 
\includegraphics[width=40mm]{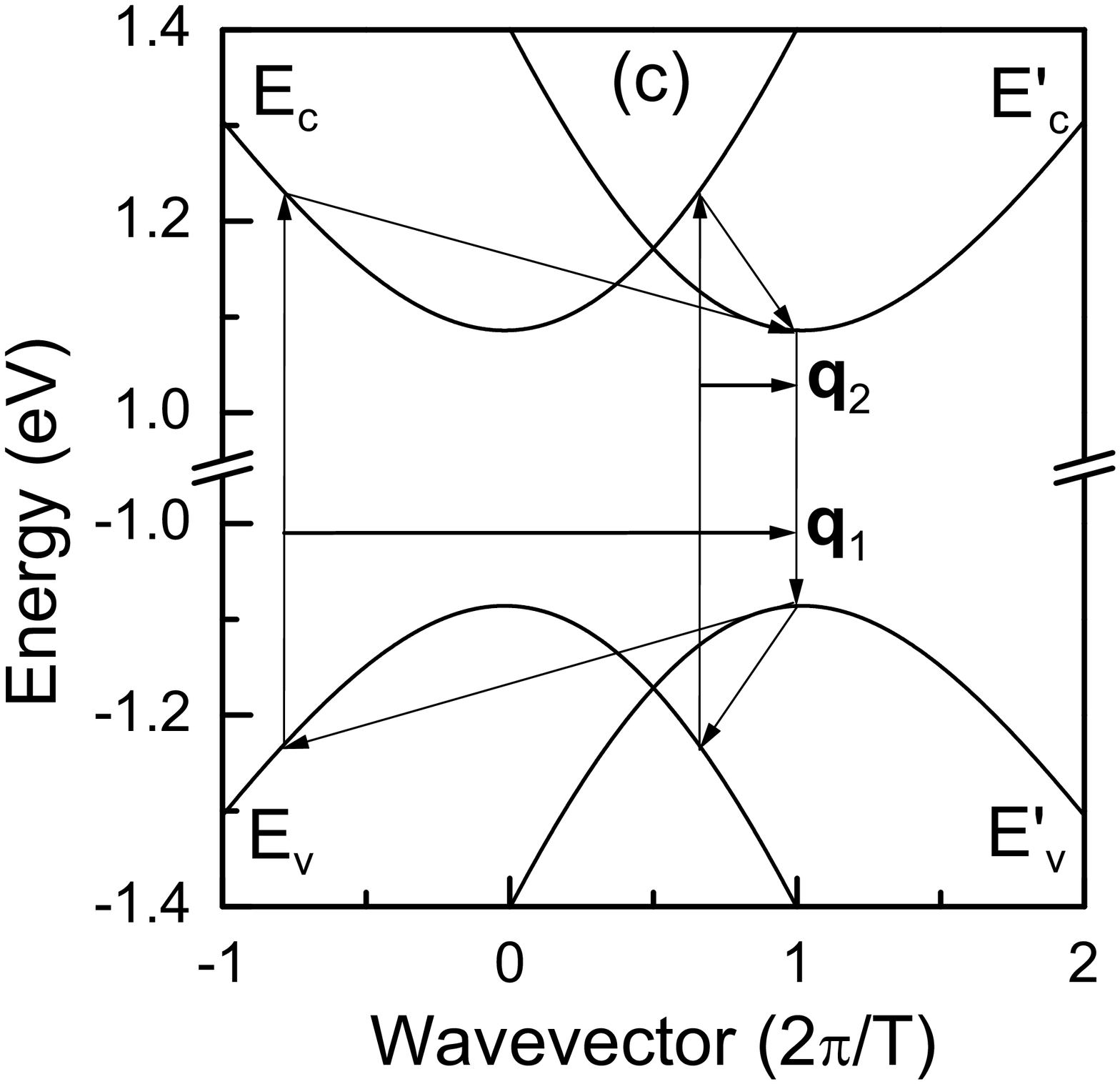} 
\includegraphics[width=40mm]{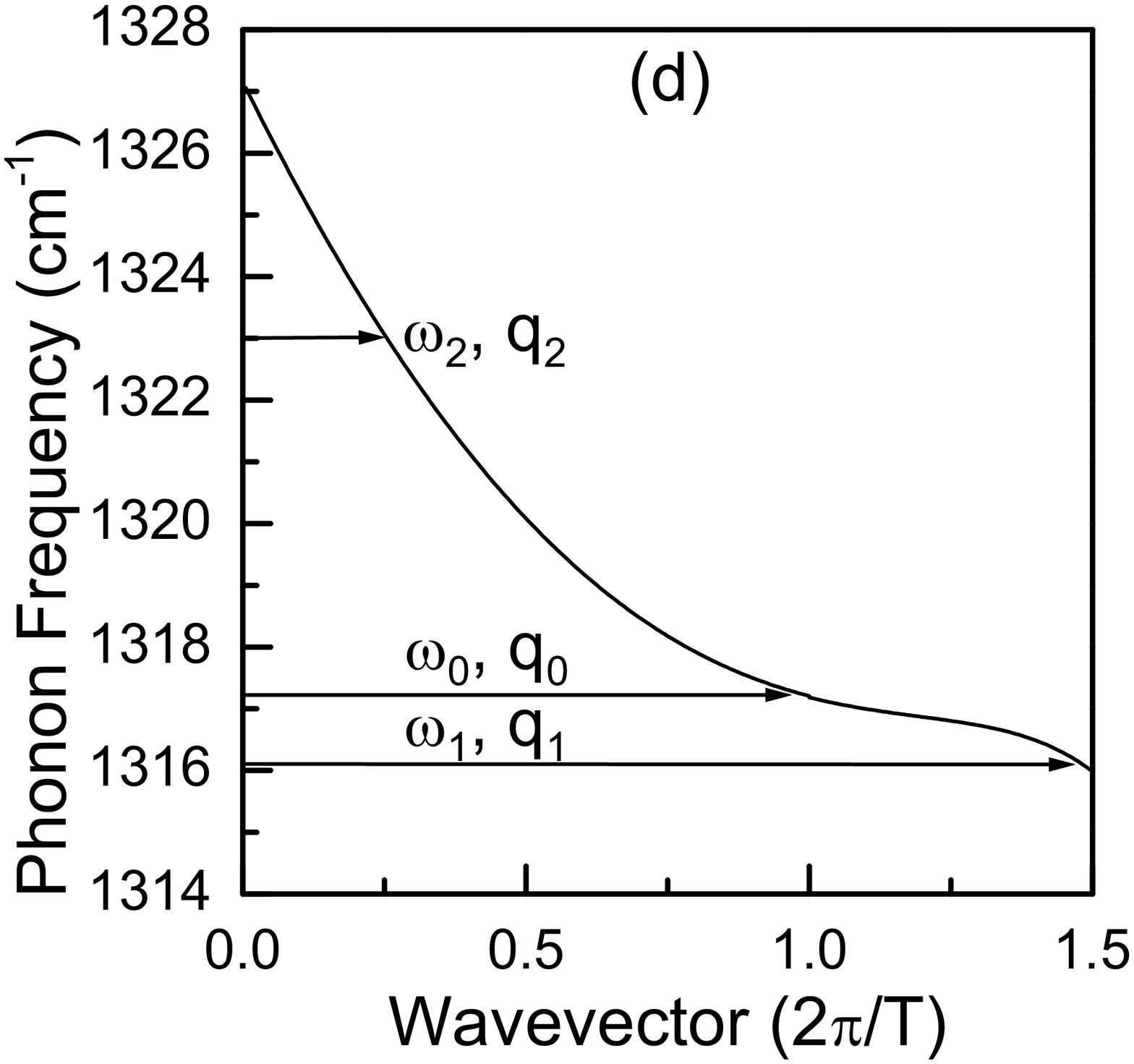} 
\caption{Part of the electronic band structure of nanotube $(6,5)$ in the extended zone representation with characteristic diagrams for \textit{ee} and \textit{eh} two-phonon processes with largest (a) single resonance, (b) double resonance, and (c) triple resonance contribution to the 2D band. The diagrams are closed polygons of arrows, denoting the virtual processes. (d) Part of the TO branch of nanotube $(6,5)$ in the extended zone representation, including phonons, relevant to the two-phonon processes in (a), (b), and (c).}
\end{figure}

The two-peaked structure of the Raman spectrum, Fig. 3a, and the three-peaked structure of the integrated intensity, Fig. 3b, can be explained with the vHs of the electronic density of states and the small denominator in Eq. 1 for small $E_{iu}$'s. In the case of mirror pairs of conduction parabolic bands and mirror pairs of valence parabolic bands with respect to $k=\pi/T$, which is characteristic for nanotubes (Fig. 1a), there are two types of scattering processes that give rise to the 2D band (see also Ref.~\onlinecite{vene11}):

(a) \textit{ee} (or  \textit{hh}) processes, with twice scattering of an electron (or a hole) by phonons with opposite momenta, for which $E_{ia}=E_L-E_c+E_v$, $E_{ib}=E_L-E_c+E^{`}_v-\hbar\omega$ (or $E_{ib}=E_L-E^{`}_c+E_v-\hbar\omega$), and $E_{ic}=E_L-E_c+E_v-2\hbar\omega$,

(b) \textit{eh} processes with scattering of an electron and a hole by two phonons with opposite momenta, for which  $E_{ia}=E_L-E_c+E_v$, $E_{ib}=E_L-E_c+E^{`}_v-\hbar\omega$ (or $E_{ib}=E_L-E^{`}_c+E_v-\hbar\omega$), and $E_{ic}=E_L-E^{`}_c+E^{`}_v-2\hbar\omega$. 

First of all, single resonances are present for processes, for which one of the $E_{iu}$'s vanishes. However, this does not yield a significant increase of the Raman intensity unless the process involves initial and final electron wavevectors at a vHs. Such electron wavevectors are connected by phonon wavevectors, close to the wavevector $q_0$ between the extrema of mirror pairs of conduction bands and mirror pairs of valence bands (Fig. 4a). Thus, the scattering phonons will have frequency, close to $\omega_0 \equiv \omega(q_0)$ (Fig. 4d) and the Raman shift of the 2D band will be equal to $2\omega_0$ (Fig. 3a). For the mentioned phonon wavevectors, $E_{ia}$, $E_{ib}$, and $E_{ic}$ turn to zero  at $E_L = E_{22}$, $E_L = E_{22} + \hbar \omega_0$, or $E_L = E_{22} + 2\hbar \omega_0$, respectively. This corresponds to the derived three-peaked structure of the integrated intensity of the 2D band (Fig. 3b). 

Double resonances are possible only for type \textit{ee}/\textit{hh} processes but the largest contribution comes from processes with initial or final electron wavevector at a vHs. There are two phonon wavevectors for such processes, $q_1$ and $q_2$ (Fig. 4b). These phonons generally have different frequencies $\omega_1 \equiv \omega(q_1)$ and $\omega_2 \equiv \omega(q_2)$ (Fig. 4d) and give rise to two peaks of the 2D band with Raman shifts of $2\omega_1$ and $2\omega_2$ (Fig. 2a). In this case, $E_{ia}$ and $E_{ib}$ simultaneously turn to zero at $E_L = E_{22} + 2\hbar \omega_{1}$ or $E_L = E_{22} + 2\hbar \omega_{2}$. These processes will contribute to a double peak of the integrated intensity at $E_L = E_{22} + 2\hbar \omega_1$ and $E_L = E_{22} + 2\hbar \omega_2$. However, because $\omega_2 - \omega_1$ is much smaller than the electronic linewidth, the two peaks are likely to be observed as a single one at $E_L = E_{22} + 2\hbar \omega$ where  $\omega \in [\omega_1,\omega_2]$. 

Finally, triple resonances are possible only for type \textit{eh} processes. In this case, the intensity is largest for initial or final electronic states at a vHs. Similarly to the case of double resonance, there are again two phonon wavevectors, $q_1$ and $q_2$, satisfying this condition (Fig. 4c). The corresponding phonon frequencies $\omega_1$ and $\omega_2$ are generally different (Fig. 4d) and give rise to two peaks of the 2D band with Raman shifts of $2\omega_1$ and $2\omega_2$ (Fig. 3a). All three factors $E_{iu}$ in the denominator of Eq. 1 vanish simultaneously at $E_L = E_{22} + 2\hbar \omega_1$ or $E_L = E_{22} + 2\hbar \omega_2$. Triple resonance processes will then contribute to the peak of the integrated intensity at $E_L = E_{22} + 2\hbar \omega$, where $\omega \in [\omega_1,\omega_2]$.

The provided description of the contributions of the various scattering processes to the 2D band allows analyzing the complex shape of this band. With increasing $E_L$ from below $E_{22}$, the 2D band is initially a single-peak one, mainly due to single resonance processes. At higher $E_L$, the 2D band shape evolves into a two-peak one, due to the dominant contributions of double and triple resonance processes; the two peaks of the 2D band are red-shifted and blue-shifted, depending on the behavior of the TO phonon branch. We note that for such $E_L$ there will also be a contribution to the 2D band from single resonances due to TO phonons with frequency $\omega_0$, but it is much smaller than those from double and triple resonances and cannot to be resolved. The predicted two-peaked 2D band corresponds well to the recent experimental data.\cite{mour17} 

The evolution of the 2D band of any nanotube with laser excitation is expected to show similar behavior as that for nanotube $(6,5)$. Namely, the 2D band will split into a two-peaked structure with increasing $E_L$. The red shift and blue shift of the constituent peaks will depend on the TO branch and the optical transition of the nanotube. For small enough slope of the TO branch, splitting of the 2D band may not be observed at all.

The characteristic three-peak shape of the integrated intensity vs laser excitation is quite different from that in graphene because the former is intrinsically connected to the characteristic vHs in nanotubes, which are not present in graphene. Also, unlike graphene, it is difficult to define dispersion rate of the 2D band, because of the complex shape of this band. However, dispersion rate can be associated with each of the peaks of the 2D band and can be deduced from the slope of the part of the TO branch, which is relevant to the 2D band. 

The behavior of the remaining two-phonon bands is similar to that of the 2D band. Namely, the Raman bands have significant intensity for laser excitation roughly in the range between the optical transition $E_{ii}$ and $E_{ii} + 2\hbar\omega$. For $E_L$ close to $E_{ii}$, single resonance processes give rise to a non-dispersive two-phonon band. With increasing $E_L$, double and triple resonance processes become dominant and the two-phonon band splits into two dispersive components.

\section{Conclusions}

We presented a computational approach to the calculation of the two-phonon bands of carbon nanotubes, based on a symmetry-adapted non-orthogonal tight-binding model. As a case study, we considered the narrow nanotube $(6,5)$ and analyzed the evolution of the 2D Raman band vs laser excitation. We found that this band splits into two dispersive peaks with increasing laser excitation energy. Such behavior is expected for any two-phonon band of any nanotube with dispersion rate of the dispersive peaks, depending on the electronic and phonon dispersion, and laser excitation. The adopted symmetry-adapted approach significantly reduces the computational efforts in comparison with the approach, based only on the translational symmetry of the nanotubes, and allows one to derive the two-phonon Raman bands of any observable carbon nanotube.

\acknowledgments

V.N.P. acknowledges financial support by the National Science Fund of Bulgaria under grant DN18/9-11.12.2017.
 

%

\end{document}